\documentclass[prl,reprint,superscriptaddress]{revtex4-1}
\usepackage{graphicx}
\usepackage{amsmath}
\usepackage{amssymb}
\usepackage{epstopdf}
\usepackage{xcolor}
\usepackage{subfigure}
\usepackage{blindtext}

\graphicspath{{./FIGURES/}}

\begin{document}

\title{Tunable Polarization-Induced Fano Resonances in Stacked Wire-Grid Metasurfaces}

\author{Xavier Romain}
\email{email: x.romain@lancaster.ac.uk}
\affiliation{D\'epartement Optique - Institut FEMTO-ST UMR 6174, Universit\' e Bourgogne Franche-Comt\'e - CNRS, 25030 Besan\c con, France}
\affiliation{Department of Engineering, Lancaster University, Bailrigg, Lancaster LA1 4YW, UK}
\author{Riccardo Degl'Innocenti}
\affiliation{Department of Engineering, Lancaster University, Bailrigg, Lancaster LA1 4YW, UK}
\author{Fadi I. Baida}
\affiliation{D\'epartement Optique - Institut FEMTO-ST UMR 6174, Universit\' e Bourgogne Franche-Comt\'e - CNRS, 25030 Besan\c con, France}
\author{Philippe Boyer}
\affiliation{D\'epartement Optique - Institut FEMTO-ST UMR 6174, Universit\' e Bourgogne Franche-Comt\'e - CNRS, 25030 Besan\c con, France}

\date{\today}

\begin{abstract}
Stacked metasurfaces are being investigated in light of exploring exotic optical effects that cannot be achieved with single-layered metasurfaces. In this Letter, we theoretically demonstrate that stacks of metallic wire-grid metasurfaces possessing specific polarization properties have the ability to induce tunable Fano resonances. The developed original model - combining a circulating field approach together with an extended Jones formalism - reveals the underlying principle that gives rise to the polarization-induced Fano resonances. The theoretical frame is validated in an experimental proof of concept using commercially available wire-grids and a terahertz time domain spectrometer. This unexplored possibility opens an alternative path to the realization and  control of Fano resonances by using stacked metallic metasurfaces. Furthermore, these findings suggest that the polarization can be used as an additional degree of freedom for the design of optical resonators with enhanced and tunable properties.
\end{abstract}


\maketitle 

%\section{Introduction}

Fano resonances \cite{fano_effects_1961,limonov_fano_2017} currently draw much attention because of their remarkable and unique potential for applications such as sensing with high quality factor \cite{singh_ultrasensitive_2014}. Over the past decades, Fano resonances have been reported in a large variety of experimental configurations ranging from electromagnetic structures \cite{fan_sharp_2002,fan_analysis_2002,fan_temporal_2003,lee_fano_2004} to elastic structures \cite{goffaux_evidence_2002}. Similar observations of Fano signature in electromagnetic metamaterials \cite{fedotov_sharp_2007} were made by deliberately breaking the symmetry of the metamaterial unit-cell. These results have been rapidly followed by extensive studies on the coupling of trapped modes in metamaterials in order to obtain Fano resonances \cite{singh_coupling_2009}. More recently, a vast literature focused on the excitation of tunable and/or multiple Fano lineshapes for increased functionalities \cite{yang_characteristics_2018,fu_independently_2018,zhou_tunable_2018}. At the same time, metallic metamaterials have rised in popularity because of the wide diversity of physical effects which they can exhibit such as extraordinary optical transmission \cite{ebbesen_extraordinary_1998}, negative refraction \cite{smith_metamaterials_2004} or perfect absorption \cite{landy_perfect_2008}. However, the performances associated to these physical effects may be limited by the intrinsic physical properties of the unit-cell, or might require complex designs and fabrications. Stacked metasurfaces are currently proposed as an alternative way to achieve complex functionalities \cite{cheng_emergent_2015}. Besides, the interaction between metasurfaces brings further degree of freedom and enables additional effects \cite{liu_three-dimensional_2008}. For example, stacked structures are currently proposed to efficiently manipulate the polarization of light \cite{boutria_tunable_2012,pfeiffer_high_2014,fan_freely_2015,romain_spectrally_2017}.

In this Letter, we demonstrate that stacked metallic metasurfaces offer the possibility to realize tunable Fano resonances that are induced by the specific polarization properties of the constitutive metasurfaces. The polarization-induced Fano resonance mechanism is theoretically analyzed with an original cavity model which combines a circulating field approach \cite{ismail_fabry-perot_2016} and an extended Jones formalism \cite{boyer_jones_2014}. The theoretical study is experimentally supported by demonstrating tunable polarization-induced Fano resonances in the THz regime by using a THz time domain spectrometer and implementing Metallic Wire-Grid Metasurfaces (MWGMs).

\begin{figure}[b!]
\centering
\includegraphics[width=.48\textwidth]{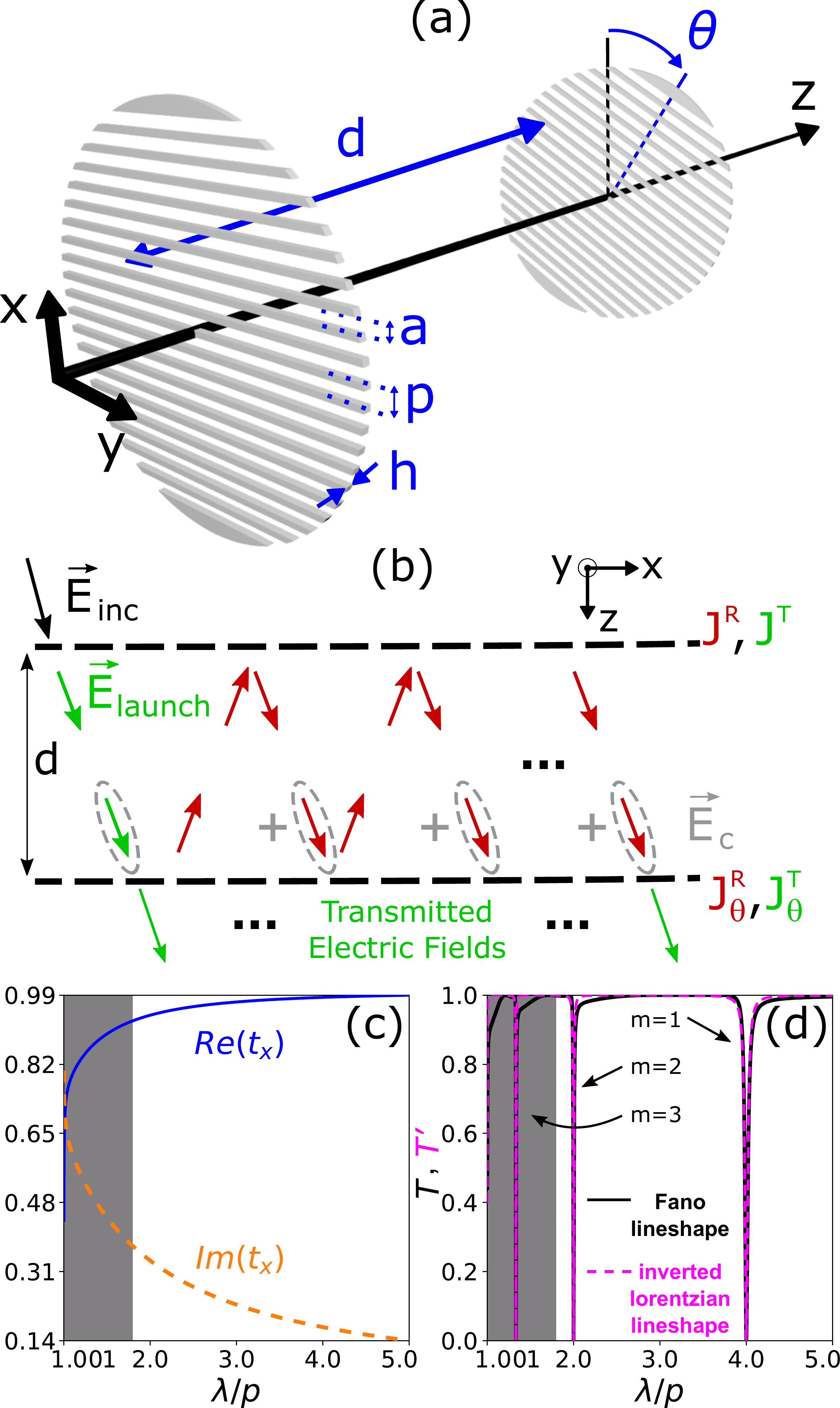}
\caption{\textbf{(a)} 3D illustration of two MWGMs stacked along the z-axis where the second MWGM is rotated by an angle $\theta$. \textbf{(b)} Principle of the FP-like cavity formed by two MWGMs characterized by their Jones matrices $J^T$, $J^R$ and $J^T_{\theta}$, $J^R_{\theta}$ respectively. $\vec{E}_{launch}$ is the initial electric field entering the cavity, and $\vec{E}_{c}$ is the steady state forward circulating field in the FP-like cavity. The arrows are deliberately tilted to clearly represent the round trips in the FP-like cavity. \textbf{(c)} Simulated spectra of $t_x$, the transmission coefficient along the x-axis for one MWGM. The real and imaginary parts of $t_x$ are depicted in solid blue line and dashed orange line respectively, and computed with $a/p=0.9$ and $h/p=0.1$. The greyed area specifies the spectral region ($\lambda/p\leq 1.8$) for which the model has lower accuracy. \textbf{(d)} Simulated transmission spectra of the FP-like cavity with $\theta=10^{\circ}$ and $d/p=2.0$. The integer $m$ denotes the FP harmonic order where the Fano resonances occurs. The black curve shows the spectrum of the transmission $T$ computed  with the numerical values extracted from the monomode modal method. The dashed pink curve gives the spectrum of the transmission $T'$ computed with the special case where $t_x=1$ and $r_x=0$.}
\label{Fig:BrightDarkModes}
\end{figure}

The importance of the MWGMs linear polarization properties for exciting Fano resonances is first examined. An electromagnetic plane wave is considered to be propagating along the z-axis and falls in normal incidence on a stack of two parallel and perfectly conducting MWGMs, as shown in Fig. \ref{Fig:BrightDarkModes}\textbf{(a)}, acting as linear polarizers. The transmission and reflection of the first MWGM, identifying the x and y axes respectively, serve as a reference. The period, the thickness and the aperture width of both MWGMs are denoted by $p$, $h$ and $a$ respectively. The distance between the two MWGMs is $d$ and $\theta$ is the rotation angle of the second MWGM with respect to the x-axis. The geometrical notations of the structure are summarized in Fig. \ref{Fig:BrightDarkModes}\textbf{(a)}.

The stack of two MWGMs is assumed to form a Fabry-Perot-like (FP-like) cavity. To accurately describe the resonance and polarization properties of the FP-like cavity, a circulating field approach \cite{ismail_fabry-perot_2016} is associated to a Jones formalism \cite{boyer_jones_2014}, as illustrated in Fig. \ref{Fig:BrightDarkModes}\textbf{(b)}. A scalar model (excluding polarization properties), reporting more details on the circulating field was thoroughly investigated in \cite{ismail_fabry-perot_2016}. The steady state forward circulating field, $\vec{E}_{c}$, is given by
\begin{equation}
\vec{E}_{c} = J_{c}U\vec{E}_{launch}
\label{Eq:T_FP_Iso}
\end{equation}
where $\vec{E}_{c}$ corresponds to the infinite sum of waves incoming on the second MWGM, as highlighted by the dashed grey ellipses in Fig. \ref{Fig:BrightDarkModes}\textbf{(b)}. $J_c$ is the Jones matrix  that accounts for the infinite round-trips in the cavity and its expression is detailed later in Eq. (\ref{Eq:J_c}), Eq. (\ref{Eq:J_rt}) and Eq. (\ref{Eq:Expanded_Jrt}). The propagation operator $U=uI$ links the electric fields from the first to the second MWGM inside the FP-like cavity. The term $I$ is a $(2\times 2)$ identity matrix, $u=e^{id2\pi /\lambda}$ represents the phase shift accumulated by the electric field in half a round-trip and $\lambda$ denotes the wavelength. The electric field $\vec{E}_{launch}$, is the initial field launched in the cavity and its expression is
\begin{equation}
\vec{E}_{launch} = J^T\vec{E}_{inc}
\end{equation}
where
\begin{equation}
J^T = \left ( \begin{array}{cc}
t_x & 0 \\
0 & t_y
\end{array} \right ) \textrm{  and  } J^R = \left ( \begin{array}{cc}
r_x & 0 \\
0 & r_y
\end{array} \right )
\end{equation}
are the transmission and reflection Jones matrices of the first MWGM where $J^{R}$ is mentioned for completeness. The terms $t_{x}$, $t_y$ and $r_{x}$, $r_y$ are respectively the transmission and reflection coefficients along the $x$ and $y$ axes for one MWGM. The polarization-induced effect can be inferred from the expression of $J_c$ which is written as
\begin{equation}
J_{c}=\left [ I - U^2J^RJ^R_{\theta}\right ] ^{-1}
\label{Eq:J_c}
\end{equation} where $J^R_{\theta}= R(\theta)J^RR(-\theta)$ is the reflection Jones matrix of the second MWGM and $R(\theta)$ is the rotation matrix.
The matrix $J_{c}$ can be reformulated as
\begin{equation}
J_{c} = \left ( \begin{array}{cc}
J_{c}^{xx} & J_{c}^{xy} \\
J_{c}^{yx}& J_{c}^{yy}
\end{array} \right )
\label{Eq:J_rt}
\end{equation}
which can be expanded into
\begin{equation}
\medmuskip=-0.4mu
\thinmuskip=0.1mu
\thickmuskip=0.1mu
J_{c} = \dfrac{1}{D}\left ( \begin{array}{cc}
1 - u^2r_y(s^2r_x+c^2r_y) & u^2cs(r_x - r_y)r_x \\
u^2cs(r_x - r_y)r_y & 1 - u^2r_x(c^2r_x+s^2r_y)
\end{array} \right )
\label{Eq:Expanded_Jrt}
\end{equation}
where $D = det(I - U^2J^RJ^R_{\theta})$, $c=cos\theta$ and $s=sin\theta$. For this complex general case given by Eq. (\ref{Eq:Expanded_Jrt}), it is fundamental to remark that the coupling terms $J_{c}^{yx}$ and $J_{c}^{xy}$ are vanishing if $r_x = r_y$ - as it would be the case with polarization insensitive mirrors. In other words, the polarization properties of the MWGMs brings additional polarization coupling effects - inducing Fano resonances - that are not achievable in classical FP cavities. Other works using matrix formalism have analyzed anisotropic FP resonators \cite{doyle_properties_1965,mamaev_nonorthogonal_2011} or chiral FP interferometers \cite{silverman_interferometric_1994,timofeev_geometric_2015}. None of these studies, however, considered mirrors with polarization dependency, they rather employed classical FP cavities filled with anisotropic or optically active media. The linear polarization dependency of the MWGMs ($r_x \neq r_y$) is the key difference that permits the excitation of Fano resonances.

The transmission of the optical arrangement of Fig. \ref{Fig:BrightDarkModes}\textbf{(a)} is numerically simulated with $a/p=0.9$, $h/p=0.1$, $d/p=2.0$ and $\theta=10^{\circ}$. The transmission properties of one MWGM are first investigated. The spectral range of interest is $\lambda/p \geq 2$ which ensures that only the fundamental transverse electromagnetic guided mode is excited in the MWGMs. Note that the fundamental guided mode is polarized along the periodicity axis, i.e the x-axis. Using a  monomode modal method \cite{boyer_jones_2014}, the coefficient $t_x$ and $r_x$ are numerically calculated by Airy-like expressions. The perfectly conducting metal hypothesis imposes that the two other coefficients $t_y$ and $r_y$ are $0$ and $-1$ respectively. This further implies that the electric field $\vec{E}_{launch}$ is polarized along the x-axis. Figure \ref{Fig:BrightDarkModes}\textbf{(c)} represents the spectra of the real and imaginary parts of $t_x$ in solid blue line and dashed orange line respectively. The transmission $T$ of the FP-like cavity is defined as
\begin{equation}
T = |J^{T,xx}_{FP}|^2+|J^{T,yx}_{FP}|^2
\label{Eq:Transmission_T}
\end{equation}
with
\begin{equation}
J^T_{FP} = J^T_{\theta}J_{c}UJ^T
\label{Eq:JTFP}
\end{equation}
where $J^T_{FP} $ is the transmission Jones matrix of the FP-like cavity and $J^T_{\theta}=R(\theta)J^TR(-\theta)$ is the transmission Jones matrix of the second MWGM. Note that the transmission results achieved by this formalism are strictly identical to the transmission values computed by the S-matrix propagation algorithm \cite{romain_extended_2016}.  Figure \ref{Fig:BrightDarkModes}\textbf{(d)} yields the transmission spectrum of the cavity in solid black line computed for $d/p=2.0$ and $\theta=10^{\circ}$. Clear Fano-like resonances appear at the FP resonance condition expressed as
\begin{equation}
\lambda=\dfrac{2d}{m}
\label{Eq:FP_cond}
\end{equation}
where $m$ denotes the FP harmonic order with $m \in \mathbb{N}$. It is worth emphasizing that the Fano dip asymmetry increases as $\lambda$ decreases. However, the greyed area for which $\lambda/p \leq 1.8$ is the region where higher order modes are guided in the MWGMs and the monomode modal method reaches a lower computation accuracy for $t_x$.

To further explain the origin of these Fano resonances, the lower FP harmonic $m=1$ is analyzed in more details. At $\lambda/p=4.0$, the coefficients are $t_x=0.985 + 0.169i$ and $r_x =0.475\times 10^{-2} +  0.028i$. For explanatory purposes, it is now assumed that $t_x=1$ and $r_x=0$. In this ideal and almost realistic scenario the matrix $J_c$ reduces to 
\begin{equation}
J_{c} = \left (\begin{array}{cc}
1 & 0 \\
-\dfrac{u^2cs}{1-u^2c^2} & \dfrac{1}{1-u^2c^2}
\end{array} \right ).
\label{Eq:J_C_Res}
\end{equation}
This specific peculiar case is interesting from a theoretical perspective as it helps elucidating the physical mechanism which gives rise to Fano resonances. To highlight the core principle responsible for the polarization-induced Fano resonance , it is now essential to distinguish two cases:
\begin{itemize}
\item[1)] When the MWGMs are aligned ($\theta=0^{\circ}$), the coupling terms are $J_{c}^{yx}=J_{c}^{xy}=0$. This configuration yields $J_{c}^{yy} = 1/(1-u^2)$ corresponding to the electric field Airy distribution of a perfect y-polarized FP resonance referred to as FP$_y$. At the same time, Eq. (\ref{Eq:JTFP}) becomes \begin{equation}
J^T_{FP} = J^TJ_cUJ^T = \left (\begin{array}{cc}
u & 0 \\
0 & 0
\end{array}\right )
\label{Eq:single_pass_prop}
\end{equation} which represents an x-polarized single pass propagation operator. The x-polarized field $\vec{E}_{launch}$ does not couple to the FP$_y$ resonance and simply propagates through the structure. Therefore, the FP$_{y}$ resonance is not excited or, in other word, it is "trapped" in the FP-like cavity.
\item[2)] When the MWGMs are not aligned ($\theta \neq 0 ^{\circ}$), the coupling of $\vec{E}_{launch}$ to the FP$_y$ resonance occurs via the term $J_{c}^{yx}$. As it is numerically demonstrated in Fig. \ref{Fig:BrightDarkModes}\textbf{(d)}, such coupling induces a Fano resonance according to the FP resonance condition given by Eq. (\ref{Eq:FP_cond}). The single pass x-polarized electric field and the "trapped" FP$_{y}$ resonance might be regarded as an analogue of bright and dark modes respectively, that are exploited to excite Fano resonances \cite{singh_coupling_2009}.
\end{itemize}
Finally, for this special case where $t_x=1$ and $r_x=0$, the transmission $T'$ through the FP-like cavity is
\begin{equation}
T' = \left |\dfrac{u(1 - u^2)}{1 - u^2\cos^2\theta}\right |^2 \cos^2\theta
\end{equation}
which is consistent with the transmission expression reported in ref.  \cite{romain_extended_2016}. The FP resonance condition, as stated by Eq. (\ref{Eq:FP_cond}), corresponds to the case where $u=1$ and therefore leads to $T' = 0, \ \forall  \ \theta \neq 0 \pmod {\pi}$. The expression of $T'$ at the FP resonance condition produces inverted Lorentzian line shapes, as depicted by the dashed pink curve in Fig. \ref{Fig:BrightDarkModes}\textbf{(d)} and as predicted by Eq. (17) of ref. \cite{fan_temporal_2003}. The results presented in Fig. \ref{Fig:BrightDarkModes}\textbf{(d)} are also compatible with ref. \cite{fan_temporal_2003}. Indeed, $T \rightarrow T'$ when the order $m$ decreases since $t_x \rightarrow 1$ when $\lambda \rightarrow \infty$, as shown in Fig. \ref{Fig:BrightDarkModes}\textbf{(c)}. It is important to note that, in Eq. (\ref{Eq:J_C_Res}), the coupling term $J_{c}^{yx}$ can be controlled by $\theta$. It further suggests that the spectral width of the polarization-induced Fano resonances can be tuned by acting on $\theta$, as previously observed in ref.  \cite{romain_extended_2016}.
\begin{figure}[b!]
\centering
\includegraphics[width=0.48\textwidth]{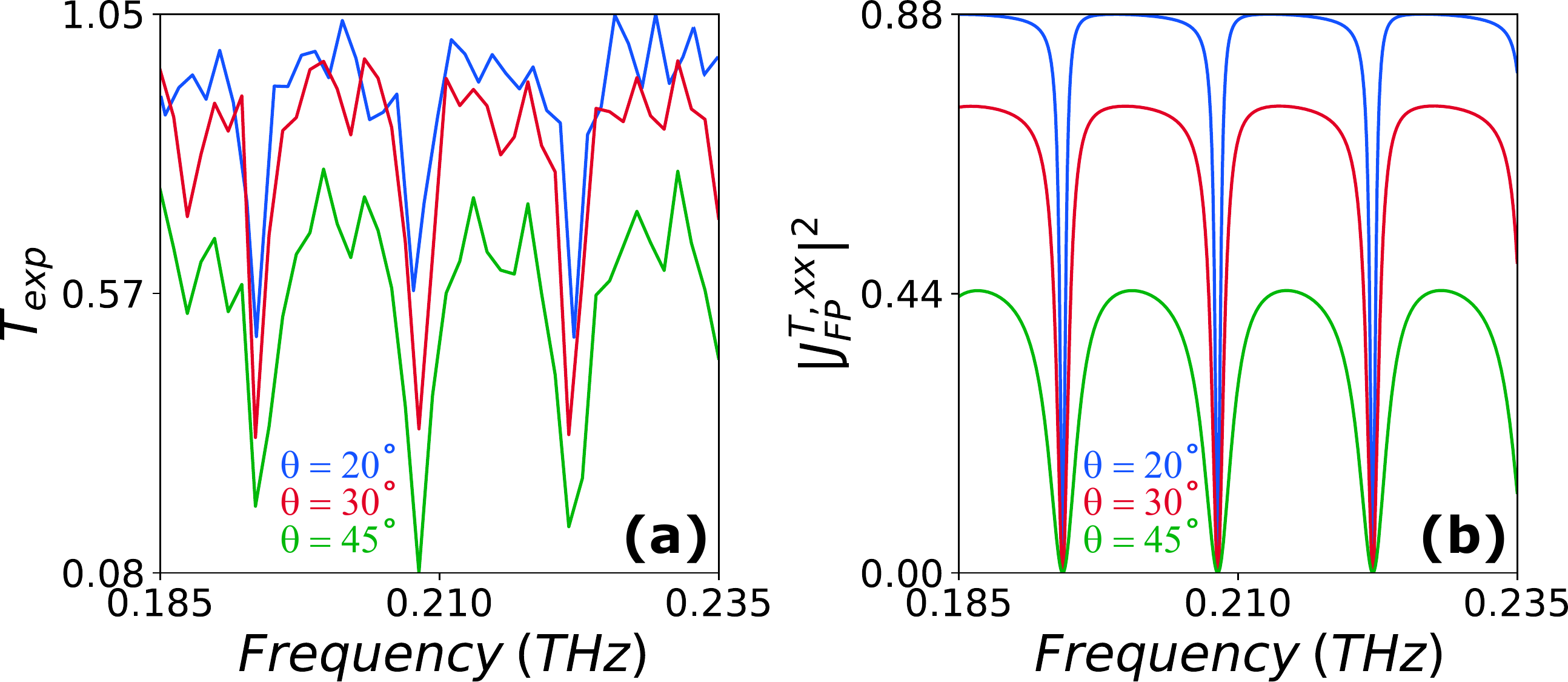}
\caption{\textbf{(a)} Measured and \textbf{(b)} computed transmission spectra for $p = 35 \ \mu m, \ a = 25 \ \mu m, \ h = 10 \ \mu m$ and $d=1,08 \ cm$.}
\label{Fig:Exp}
\end{figure}

The polarization-induced Fano resonance effect and its spectral tunability are experimentally demonstrated in the THz region. The THz range is particularly relevant and suited to verify the developed theoretical model as metals are close to be considered as perfect electric conductors at these frequencies. A pair of commercially available MWGMs from PureWavePolarizer \cite{PureWavePolarizer} is placed in the same configuration as previously depicted in Fig. \ref{Fig:BrightDarkModes}\textbf{(a)}. The geometrical parameters are $p = 35 \ \mu m, \ a=25 \ \mu m, \ h = 10 \ \mu m,$ and $d=1,08 \ cm$. The two MWGMs are inserted into a Menlo Systems Tera K15 THz time domain spectrometer based on photoconductive antennas \cite{TeraK15}. The maximal spectral resolution of the spectrometer is 1.2 GHz in the range of 0.1 THz to 3.5 THz. The first MWGM's transmission axis is aligned to the polarization direction of the photoconductive emitter. A set of transmission measurements, denoted by $T(\theta)$, are acquired at different relative angles of the second MWGM, i.e. for $\theta = \ 20^{\circ}, \ 30^{\circ}$ and $45^{\circ}$. The transmission $T_0$, measured at $\theta= 0^{\circ}$, serves as a reference for the normalized transmission $T_{exp}$ which is defined as
\begin{equation}
T_{exp} = \frac{T(\theta)}{T_0}.
\end{equation}
Figure \ref{Fig:Exp}\textbf{(a)} represents the normalized transmission spectra for the different angles $\theta$ and Fig. \ref{Fig:Exp}\textbf{(b)} depicts the computed counterpart using the FP-like cavity model. In order to be commensurate with the detector's polarization sensitivity, the numerical transmission is given by $|J_{FP}^{T,xx}|^2$. The measured and simulated results reported in Fig. \ref{Fig:Exp}\textbf{(a)} and \textbf{(b)} respectively, exhibit a remarkable agreement only limited by the available spectrometer resolution and by the signal to noise ratio.

In conclusion, stacked MWGMs with linear polarization dependency have been theoretically investigated and experimentally confirmed to be an efficient approach to induce and control Fano resonances. The experimental measurements were performed in the THz spectral range showing very good agreement with the predicted trends. It is worth stressing the versatility and generalization of this alternative way to realize Fano resonances by using FP-like cavities. This different approach does not require  metasurfaces that are specially designed with complex unit-cells or custom-made materials. Rather, the polarization-induced Fano resonance addresses fundamental concepts of optics as it essentially relies on simple polarization properties and basic resonance effects. This principle could therefore be extended to other metasurfaces made of unit-cells featuring the same linear polarization properties \cite{boyer_jones_2014}. Likewise, the polarization-induced Fano resonances could be scaled to other frequency ranges and they are currently being investigated in the visible region where the metal absorption has to be considered. The polarization-induced Fano resonances could be implemented in actively tunable optoelectronic platforms \cite{kindness_active_2018} as an additional tool to achieve enhanced performances or to offer more functionalities.

\bibliography{./BIBLIOGRAPHY/bibliography,./BIBLIOGRAPHY/ADD,./BIBLIOGRAPHY/ADD_190122,./BIBLIOGRAPHY/ADD_190124,./BIBLIOGRAPHY/ADD_190426,./BIBLIOGRAPHY/ADD_200225,./BIBLIOGRAPHY/ADD_200401}
\end{document}